\documentclass[11pt,twoside]{article}
\usepackage{asp2010}

\resetcounters

\begin{document}

\title{The substellar mass function in the central region of the open
  cluster Praesepe from deep LBT observations} \author{W. Wang$^{1}$,
  S.  Boudreault$^{1,2,3}$, B. Goldman$^{1}$, Th. Henning$^{1}$, J. A.
  Caballero$^{4}$ and C. A. L. Bailer-Jones$^{1}$ \affil{$^1$
    Max-Planck-Institut f\"ur Astronomie, K\"onigstuhl 17, D-69117
    Heidelberg, Germany} \affil{$^2$Mullard Space Science Laboratory,
    University College London, Holmbury St Mary, Dorking, Surrey, RH5
    6NT, United Kingdom} \affil{$^3$Visiting Astronomer at the
    Department of Physics and Astronomy, State University of New York,
    Stony Brook, NY 11794-3800, USA} \affil{$^4$ Centro de
    Astrobiolog\'{\i}a (CSIC-INTA), Carretera de Ajalvir km~4, 28850
    Torrej\'on de Ardoz, Madrid, Spain}}

\begin{abstract}
  Studies of the mass function (MF) of open clusters of different ages
  allow us to probe the efficiency with which brown dwarfs (BDs) are
  evaporated from clusters to populate the field. Surveys in old
  clusters (age\,$\gtrsim$\,100\,Myr) do not suffer so severely from
  several problems encountered in young clusters, such as
  intra-cluster extinction and large uncertainties in BD 
  models. Here we present the results of a deep photometric survey to
  study the MF of the old open cluster Praesepe (age
  590$^{+150}_{-120}$\,Myr and distance 190$^{+6.0}_{-5.8}$\,pc), down
  to a 5$\sigma$ detection limit at $i\sim$25.6~mag ($\sim$40\,$M_{\rm
    Jup}$). We identify 62 cluster member candidates, of which 40 are
  substellar, from comparison with predictions from a dusty atmosphere
  model. The MF rises from the substellar boundary until
  $\sim$60\,$M_{\rm Jup}$ and then declines. This is quite different
  from the form inferred for other open clusters older than 50\,Myr,
  but seems to be similar to those found in very young open cluster,
  whose MFs peak at $\sim$10\,$M_{\rm Jup}$.  Either Praesepe really
  does have a different MF from other clusters or they had similar
  initial MFs but have differed in their dynamical evolution. We
  further have identified six foreground T dwarf candidates towards
  Praesepe, which require follow-up spectroscopy to confirm their
  nature.
\end{abstract}

\section{Introduction}

The mass functions (MFs) of stellar and substellar populations have
been determined from optical and near-infrared surveys for several
open clusters at different ages, such as the Orion Nebula Cluster,
$\sigma$~Orionis, $\rho$~Ophiuchi, Taurus, IC~348, IC~2391, M35, the
Pleiades, and the Hyades. Studies of relatively old open clusters
(age\,$>$\,100\,Myr) are important for the following two reasons in
particular: first, they allow us to study the intrinsic evolution of
basic properties of BDs, e.g.\ luminosity and effective temperature,
and to compare the evolution with structural and atmospheric models;
second, we may investigate how the BD and low-mass star populations as
a whole evolve, e.g.\ the efficiency with which BDs and low-mass stars
evaporate from clusters. Such an investigation has been carried out
for the Hyades \citep[and references therein]{bouvier2008} and for
Praesepe \citep[and references therein]{boudreault2010}.

\citet[hereafter B2010]{boudreault2010} observed a significant difference between the
MFs of Praesepe and Hyades: While the Hyades MF is observed to have a
maximum at $\sim$0.6\,M$_\odot$ \citep{bouvier2008}, the MF of
Praesepe continues to rise from 0.8\,M$_\odot$ down to 0.1\,M$_\odot$.
This is surprising, as both clusters share similar physical properties
(ages, mass, metallicity, and tidal radii). Disagreement between the
Praesepe and Hyades MFs could arise from variations in the clusters'
initial MFs, or from differences in their dynamical evolution
\citep{bastian2010}. Although different binary fractions could cause
the observed (system) MFs to differ, there is no clear evidence for
varying binary fractions from measurements published in the literature
(B2010). 

\section{\label{obs-data-calib} Observations and analysis}


The Large Binocular Cameras (LBCs) are two wide-field, high-throughput imaging
cameras, namely Blue (LBCB) and Red (LBCR), located at the prime focus stations
of the Large Binocular Telescope (LBT). Each LBC has a wide field of view
(23'$\times$23'), with four CCD detectors of 2048$\times$4608 pixels each,
providing images with a sampling of 0.23\arcsec/pixel. The optical design and
detectors of the two cameras are optimized for different wavelength ranges: one
for ultraviolet-blue wavelengths (including the Bessel $U$, $B$, $V$ and Sloan
$g$ and $r$ bands), and one for the red-infrared bands (including the Sloan
$i$, $z$ and Fan $Y$ bands).  In the full binocular configuration, both cameras
are available simultaneously, and both point in the same direction of the sky,
thus doubling the net efficiency of the LBT. The survey was carried out with
the $r$ filter using LBC-blue and the $izY$ filters using LBC-red, covering
the central 0.59~deg$^2$ area of Praesepe.

The standard data reduction steps for the LBT data were performed using the IDL
astronomy package and IRAF. An astrometric solution was achieved using the
Sloan Digital Sky Survey (SDSS) catalogue as a reference. The root mean square
accuracy of our astrometric solution is 0.10-0.15\,arcsec. To correct for Earth
atmospheric absorption on the photometry, we calibrated the infrared data using
the $r$, $i$ and $z$ band values of SDSS objects which were observed in the
science fields. In order to calibrate our $Y$ band photometry, we used our LBT
$i$ and $z$ photometry and the $Y$ band photometry from the United Kingdom
Infrared Telescope Infrared Deep Sky Survey (UKIDSS).

\section{\label{selection} Candidate Selection Procedure and mass
  determination}

The candidate selection procedure and the mass determination
introduced by \citet{boudreault2009} and B2010 were
adopted in the present work. We use the evolutionary tracks from
\citep{chabrier2000} and the atmosphere models from \citep{allard01}
assuming a dusty atmosphere (the AMES-Dusty model), to compute an
isochrone for Praesepe using an age of 590$^{+150}_{-120}$\,Myr, a
distance of 190$^{+6.0}_{-5.8}$\,pc, a solar metallicity and we
neglecting the reddening.

\subsection{\label{selection-1st-2nd} Candidate Selection using
  colour-magnitude and colour-colour diagrams}

\begin{figure}[!ht]
  \plotfiddle{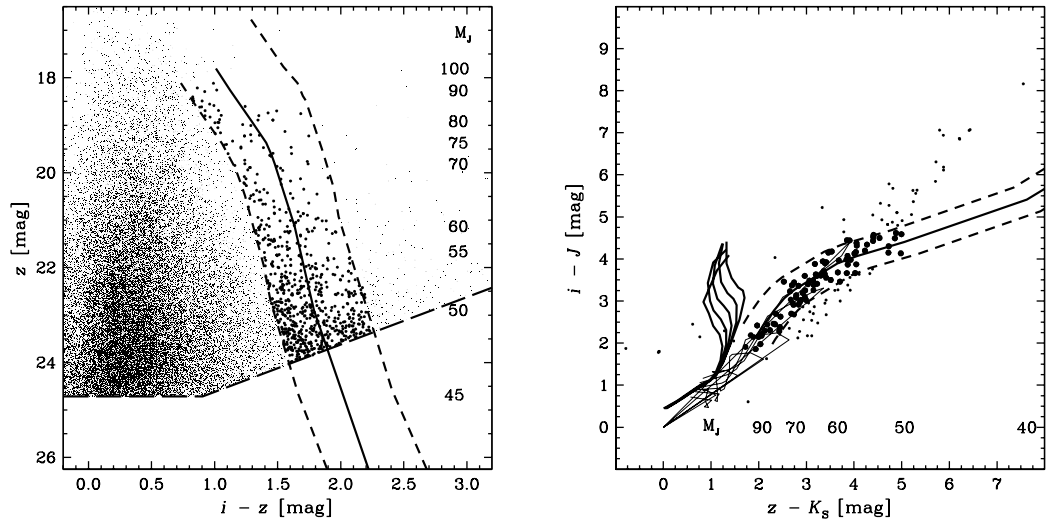}{6.5cm}{0}{70.}{70.}{-225}{-100}
  \caption{
\label{fig:cmd-iz} Colour-magnitude diagram (\textit{Left}) and colour-colour
diagram (\textit{Right}) used for the first and second selection procedures.
Solid lines are the isochrones computed from an evolutionary model with a dusty
atmosphere (AMES-Dusty). The dashed lines delimit our selection band.  The
numbers indicate the masses (in $M_{\rm Jup}$) on the model sequence.  In the
right panel, the theoretical colours of six galaxies and of red giants are
shown as thin lines and as thick lines, respectively. The six galaxies are two
starbursts, one Sab, one Sbc, and two ellipticals of 5.5 and 15\,Gyr, with
redshifts from $z$=0 to $z$=2 in steps of 0.25 (evolution not considered). We
assume that all red giants have a mass of 5\,M$_{\odot}$, 0.5 $<$ log
\textit{g} $<$ 2.5 and 2000\,K $<$ $T_{\it eff}$ $<$ 6000\,K.}
\end{figure}

Candidates were first selected from our CMD by keeping all objects which are no
more than 0.28\,mag redder or bluer than the isochrones in all CMDs. This
number accommodates errors in the magnitudes and uncertainties in the model
isochrones. We also include the errors from age and distance of Praesepe. We
additionally include objects brighter than the isochrones by 0.753\,mag in
order to include unresolved binaries. In Figure~\ref{fig:cmd-iz} (left) we show
the CMD where candidates were selected based on $z$ vs.  $i$--$z$. 

The second stage of candidate selection involves retaining just those objects
which lie within 0.28\,mag of the isochrone in the colour-colour diagram.  This
value reflects the photometric errors and uncertainties in the model
isochrones. In addition, we also include the uncertainty in the age estimation
of Praesepe. The colour-colour diagram with the selection limits is shown in
Figure \ref{fig:cmd-iz} (right), with the theoretical colours for red giants
using the atmosphere models of \citet{hauschildt1999b} and theoretical colours
of six galaxies with redshift from 0 to 2 from K. Meisenheimer et al.
(~\textit{in prep}.~) overplotted.  Neither the red giants nor the galaxies are
expected to be a significant source of contamination; most of the low redshift
galaxies can be easily rejected through visual inspection. As these are the
dominant potential contaminators, we conclude that there is no significant
contamination of non-Praesepe members in our sample (Praesepe is at a Galactic
latitude of $b=+32.5^{\circ}$).

\subsection{\label{selection-3rd} Observed magnitude vs. predicted magnitude}

\begin{figure}[!ht]
  \plotfiddle{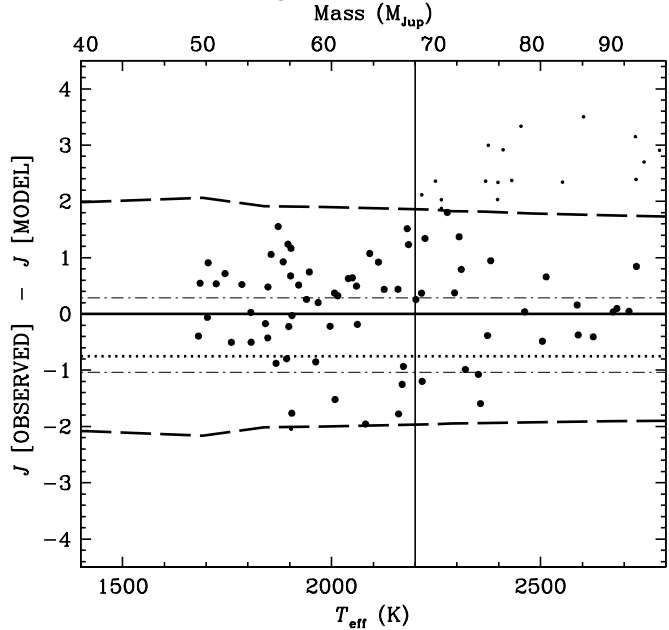}{7.5cm}{0}{60.}{60.}{-150}{-40}
  \caption{\label{fig:mj_vs_mj} Difference between the observed $J$
    magnitude and the model $J$ magnitude computed from the derived
    mass and $T_{\it eff}$, as a function of $T_{\it eff}$. The
    vertical line marks the location of L0 dwarfs. The dotted line
    (at $-0.753$\,mag) represents the offset due to the 
    presence of unresolved binaries, the dashed-dotted lines
    represent the error on the magnitude determination, and the
    long-dashed lines for the uncertainties on the age
    and distance of Praesepe. The horizontal solid line just traces
    zero.}
\end{figure}

Our determination of $T_{\rm eff}$ is based on the spectral energy
distribution of each object and is independent of the assumed
distance. The membership status of an object can therefore be assessed
by comparing its observed magnitude in a band with its magnitude
predicted from its $T_{\rm eff}$ and the Praesepe's isochrone (which
assumes a distance and an age). This selection step is only a
verification of the consistency between the physical parameters
obtained of the photometric cluster candidates with the physical
properties assume for the cluster itself when computing the
isochrones. In order to avoid removing unresolved binaries that are
real members of the cluster, we keep all objects with a computed
magnitude of up to 0.753\,mag brighter than the observed magnitude.
This selection procedure is illustrated in Fig.~\ref{fig:mj_vs_mj}.

\section{\label{results-survey} Results}

\subsection{Selected photometric candidates}

62 photometric candidates survive the selection procedures (based on isochrones
assuming dusty atmospheres). This corresponds to $\sim$\,105 objects per square
degree. Our survey saturation occurs at $\sim$ 18 mag in $z$ band, corresponding
to $\sim$100\,$M_{\rm Jup}$.  Therefore, most of the low mass candidates
discovered in previous surveys (e.g.\ \citealt{pinfield97},
\citealt{hambly1995}) saturate in our LBT images. Only a few faint BDs 
classified by \citet{pinfield97}, \citet{gonzalez-garcia2006} and by B2010 are
rediscovered by the current survey.

There are also some targets which are previously identified as cluster members
but rejected by our selection procedures or visual inspection.  For example,
ten candidates identified by B2010 are detected (not saturated) in our LBT
survey, but six of them are rejected by the $z$ vs.  $i$--$z$ CMD, because they
are bluer than the isochrones area. Another one is obviously not a point-like
source in the LBT image, and another is rejected because its observed $J$
magnitude is not consistent with its model-predicted magnitude. Only the
remaining two targets are confirmed to be cluster dwarf stars. As the current
work employed more photometric bands than B2010 did, it is not surprising that
we achieve a more conservative selection.

Most of our candidates are in the substellar regime, and other than
these ten, no other accurate photometric observations are available
from past epochs. This precludes using proper motions as a mean of
cluster membership assessment at this time.

\subsection{Contamination by non-members}

As mentioned before, the two main sources of contamination are the background
red giants and unresolved galaxies. Red giants occupy the high mass end of this
study, as seen in the $i-J$ vs. $z-K_{\it s}$ diagram.  Although some types of
galaxies share similar colours with Praesepe cluster members more massive than
60\,$M_{\rm Jup}$, such low-redshift galaxies are in general extended sources
and therefore easily rejected by our visual inspection.  Among the 74
candidates that passed our selection procedures, we identify four as galaxies
through their LBT images. Other possible sources are field L dwarfs and high
redshift quasars (for instance at $z\sim$6; \citealt{caballero2008}). However,
as such quasars have spectral energy distributions similar to mid-T dwarfs
whereas our faintest candidates have colours of early L dwarfs, and given that
they are rare (0.25 quasars at $5.5<z<6.5$ in a 0.59\,deg$^{2}$
survey, \citealt{stern2007}), the MF should not be affected by quasar
contamination.

The contamination by L dwarfs is also unimportant.  \citet{caballero2008} have
collected possible field dwarf contaminants covering spectral type from M3 to
T8 from the literature. From their Table~3, the spatial density for L dwarfs in
the solar neighbourhood is $\sim 7\times10^{-3}$\,pc$^{-3}$. Given that
Praesepe has a Galactic latitude of $+32.5$\,deg and distance of 190\,pc, the
nearby spatial density of L dwarfs Praesepe should be
$\sim\,6\times10^{-3}$\,pc$^{-3}$, assuming an exponential decrease for stellar
density perpendicular to the Galactic disk and a scale height of 500\,pc. If we
define a volume corresponding the area of our survey, and use the distance
uncertainties to the cluster as its depth, the total volume is
$\sim80$\,pc$^3$.  Therefore, we estimate that we have $\sim$0.5 L dwarf
contaminants near the cluster, which amounts to a negligible contamination of
merely 0.7\%. A similar calculation shows that we would have $\sim$4 M dwarf
contaminants, about 6\%.

We conclude that various contaminants are not important for this study
and the MF we derive for Praesepe should be accurate.

\section{\label{mf} Mass function of very low mass and substellar population of 
Praesepe}

The mass function, $\xi$(log$_{10}$M), we present here is the total
number of objects per square degree in each logarithmic mass interval
log$_{10}$M to log$_{10}$M + 0.1, Since we do not make any corrections
for binaries, we compute here a \textit{system} MF.

\begin{figure}[!ht]
  \plotfiddle{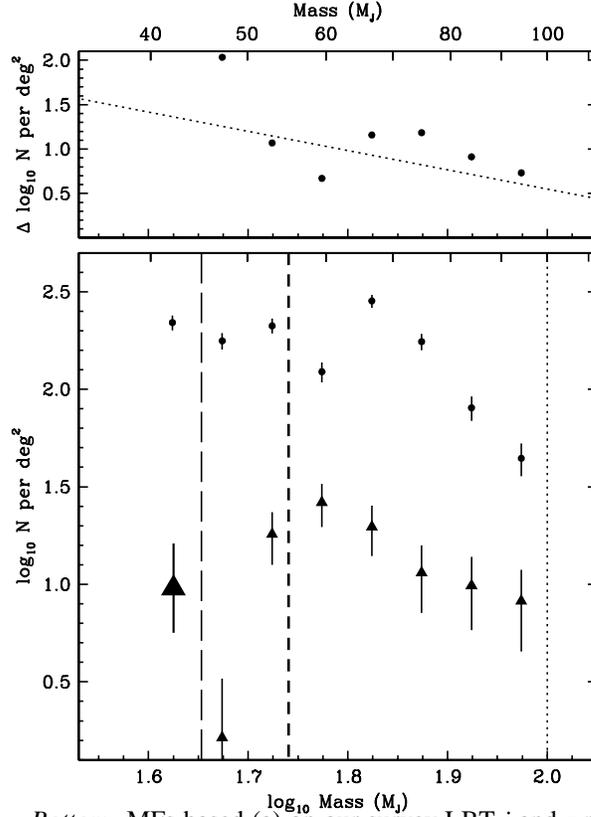}{9.5cm}{0}{60.}{60.}{-150}{-30}

  \caption{\label{fig:mf-prae-us} \textit{Bottom.} MFs based (a) on our survey
LBT $i$ and $z$ photometry (dots), and (b) also combined with the $J$ and $K_s$
photometry from B2010 (triangles). Error bars are Poissonian arising from the
number of objects in each bin, except for the last bin, for which the error bar
is mostly from the linear fit in top panel. The vertical dotted line is the
saturation mass. The vertical long and short dashed lines are the masses at the
5$\sigma$ detection limits of our optical LBT data and of the B2010 NIR data.
\textit{Top.} Difference of the number of objects in each mass bin, between
the MF computed using the LBT $iz$ data and the MF computed using
the combination of the $iz$ data from the B2010 NIR $JK_{\it s}$
data. The dotted line is a linear fit to the discrepancies.}

\end{figure}

Our optical photometry reaches lower masses than the NIR photometry that we
used. To compute the MF of Praesepe to the lowest mass bin, we first computed
a MF using only the optical $iz$ photometry. This MF is presented on
Fig.~\ref{fig:mf-prae-us} as filled dots. We computed a second MF from the list
of candidates that pass the three selections criteria which are also detected
in the survey of B2010 in the NIR $J$ and $K_{\it s}$ bands (presented on
Fig.~\ref{fig:mf-prae-us} as filled triangles). For each mass bin, we computed
the number of object removed because of adding the $J$ and $K_{\it s}$ filters
to our selection process and mass determination (plotted as a function of mass
in Fig.~\ref{fig:mf-prae-us}, top panel). We fitted a linear function to
estimate the number of objects that would be removed \textit{if} we had
additional $J$ and $K_{\it s}$ photometry down to 40--45\,$M_{\rm Jup}$.  The
corresponding extension of the MF with $izJK_{\it s}$ photometry is given as a
large triangle on Fig.~\ref{fig:mf-prae-us}.

Our derived MF of Praesepe (presented in Fig. 3) shows a rise
from 105\,$M_{\rm Jup}$ to 60\,$M_{\rm Jup}$ and then a turn-over at
$\sim$60\,$M_{\rm J}$. We note that in the second mass bin at
$\sim48\,M_{\it Jup}$, the MF is very low. This is probably because at
this mass we reach the 5$\sigma$ detection limit of the J-band of our
$\Omega$2k photometry, so we suspect this is not a real feature.
However, the turn-over at $\sim$60\,$M_{\rm Jup}$ occurs well above
the 5$\sigma$ of either $iz$ bands or $JK_{\it s}$ bands (e.g., at
$\sim$60\,$M_{\rm Jup}$, $i\sim22$, completenss$\sim100$\%) and
therefore should not be caused by incompleteness. This is the first
time one observes a clear rise in the substellar MF in old open
cluster.

\begin{figure}[!ht]
  \plotfiddle{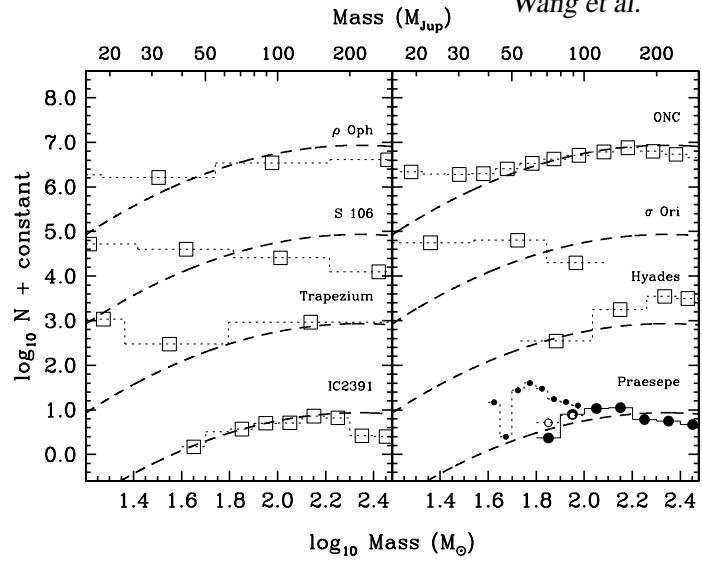}{6.0cm}{0}{60.}{60.}{-150}{-40}
  \caption{\label{fig:all-mf} Mass functions of various open cluster.
    From top to bottom: $\rho$\,Oph ($\sim$1\,Myr); S\,106
    ($\sim1$\,Myr, middle panel for example); Trapezium
    ($\sim$0.8\,Myr); IC2391 ($\sim$50\,Myr); ONC ($\sim$5\,Myr);
    $\sigma$ Ori ($\sim$3\,Myr); Hyades ($\sim$625\,Myr); Praesepe
    ($\sim$590\,Myr, from this work).  We also show the Galactic field
    star MF from \citet{chabrier2003} as a dashed line.  All the MFs
    are normalized to the lognormal fit of \citet{chabrier2003} at the
    substellar limit ($\sim$72\,$M_{\rm Jup}$).  }
\end{figure}

We collected some results for other clusters for comparisons, as
plotted in Fig.~\ref{fig:all-mf}.  This includes: IC\,2391 by
\citep{boudreault2009}, ONC by \citep{hillenbrand2000}, $\sigma$
Orionis by \citep{bihain2009} and the Hyades from \citep{bouvier2008}.
The MF of Praesepe is quite different from either IC\,2391 ( age of
$\sim50$\,Myr) or the Hyades ($\sim$625\,Myr).  Either the `dynamical
evaporation' does not have the same effect on those three clusters, or
they have a different initial mass function. Another alternative
possibility is that Praesepe has a different binary fraction. Further
studies would be necessary to clarify these points.

The continuing rise of the MF into the substellar regime which we
observe has also been observed in young clusters (as shown in
Fig.~\ref{fig:all-mf}), specifically in $\sigma$\,Orionis
\citep{bihain2009}, Trapezium \citep[turn-over at $\sim$10--20\,$M_{\rm
  Jup}$]{muench2002}, $\rho$\,Oph \citep[MF rising to
$\sim$10\,$M_{\rm J}$]{marsh2010} and in the very low
luminosity young cluster in S106, where the MF increases or at least
remains flat down to $\sim$10\,$M_{\rm Jup}$~\citep{oasa2006}.  If we
assumed a universal IMF, then it seems that the substellar MF of
Praesepe has not evolve significantly since it the cluster formed.

\section{\label{conclusions} Conclusions}

We have carried out the deepest survey to date of the open cluster Praesepe,
covering the central 0.59\,deg$^2$ in the $rizY$ bands. The survey probed a
mass range from $\sim$100 to 40$M_{\rm Jup}$ at 5$\sigma$ detection limit. Our
$iz$-bands data, combined with the $\Omega$2k NIR ($J$ \& $K_{\rm s}$) band
observations from B2010, are compared with theoretical loci of cluster members
based on a dusty atmosphere (the AMES-Dusty model), in order to clarify cluster
member candidates. Our final sample comprises 62 photometric candidates. We
estimate the contamination by field L dwarfs to be less than 1\%, and that by
galaxies and red giants also to be negligible. About two thirds of our 
candidates have theoretical masses below the Hydrogen-burning limit at
0.072\,$M_{\odot}$, and are therefore BD candidates.

The mass function we have inferred for Praesepe is consistent with that
inferred by B2010 at a mass just below the substellar boundary, but deviates by
$\sim$0.4\,dex in the next lowest mass bin, which may indicate either a
significant number of objects missing in the Boudreault et al. 2010 survey, or
a higher concentration of substellar objects in the centre of Praesepe (as the
Boudreault et al. survey is at a larger cluster radius).  The latter
possibility suggests that the dynamical evolution of very low mass stars is not
efficient in this cluster, as proposed by B2010, for explaining the discrepancy
between the Praesepe MF and Hyades MF.

The steady rise of the Praesepe MF until $\sim$60\,$M_{\rm Jup}$ was
unexpected. Such a significant peak has never been observed in any other
cluster older than 50\,Myr, but has been observed in several very young open
clusters such as $\sigma$\,Orionis or clusters in star forming regions (e.g.\
Trapezium). This suggests that the dynamical interactions in Praesepe have
very little effect on MFs, if we assume there is a universal initial MF.

The results reported here will be presented with further details in an
future publication submitted to A\&A (Wang et al. 2010,
\textit{submitted}).

\acknowledgements Some of the observations on which this work is based
were obtained during LBT programme "LBT-F08-02".

\bibliography{wang_w}

\end{document}